\documentclass[aps,prl,twocolumn,amsmath,amssymb,nofootinbib,superscriptaddress,floatfix,reprint,shortbibliography]{revtex4-2}
\usepackage{latexsym}
\usepackage{amsmath}
\usepackage{amsfonts}
\usepackage{amssymb}
\usepackage{bm}
\usepackage{color}
\usepackage{txfonts}
\usepackage{float}
\usepackage{url}
\usepackage[colorlinks=true, urlcolor=blue, linkcolor=blue, citecolor=blue, pdftex]{hyperref}
\usepackage{ulem}
\usepackage{graphicx}
\usepackage{extpfeil}
\usepackage{subfigure}
\usepackage{physics}
\usepackage{setspace}
\DeclareUnicodeCharacter{2212}{-}

\begin{document}

\newcommand{\fig}[2]{\includegraphics[width=#1]{#2}}
\newcommand{\la}{{\langle}}
\newcommand{\ra}{{\rangle}}
\newcommand{\dg}{{\dagger}}
\newcommand{\upa}{{\uparrow}}
\newcommand{\dna}{{\downarrow}}
\newcommand{\ab}{{\alpha\beta}}
\newcommand{\ias}{{i\alpha\sigma}}
\newcommand{\ibs}{{i\beta\sigma}}
\newcommand{\hH}{\hat{H}}
\newcommand{\hn}{\hat{n}}
\newcommand{\hc}{{\hat{\chi}}}
\newcommand{\hU}{{\hat{U}}}
\newcommand{\hV}{{\hat{V}}}
\newcommand{\cH}{{\mathcal{H}}}
\newcommand{\br}{{\bm{r}}}
\newcommand{\bk}{{{\bm{k}}}}
\newcommand{\bq}{{{\bm{q}}}}
\newcommand{\bp}{{{\bm{p}}}}
\def\gsim{~\rlap{$>$}{\lower 1.0ex\hbox{$\sim$}}}
\setlength{\unitlength}{1mm}
\newcommand{{\vhf}}{$\chi^\text{v}_f$}
\newcommand{{\vhd}}{$\chi^\text{v}_d$}
\newcommand{{\vpd}}{$\Delta^\text{v}_d$}
\newcommand{{\ved}}{$\epsilon^\text{v}_d$}
\newcommand{{\vved}}{$\varepsilon^\text{v}_d$}
\newcommand{\pprl}{Phys. Rev. Lett. \ }
\newcommand{\pprb}{Phys. Rev. {B}}

\title{Intra-unit-cell Josephson scanning tunneling spectroscopy of the cuprate Bi$_{1.2}$Pb$_{0.8}$Sr$_2$CaCu$_2$O$_{8+\delta}$}
\author{Fengyu Yao}
\affiliation{State Key Laboratory of Quantum Functional Materials, Department of Physics,\\Guangdong Provincial Key Laboratory of Topological Matter, \\ Guangdong Basic Research Center of Excellence for Quantum Science and, \\ Center for Advanced Light Source, Southern University of Science and Technology (SUSTech), Shenzhen 518055, China}

\author{Jiong Mei}
\affiliation{Department of Physics, The Hong Kong University of Science and Technology,\\ Clear Water Bay, Kowloon 999077, Hong Kong, China}

\author{Wenjin Mao}
\affiliation{Beijing National Laboratory for Condensed Matter Physics and,\\ Institute of Physics, Chinese Academy of Sciences, Beijing 100190, China}
\affiliation{School of Physical Sciences, University of Chinese Academy of Sciences, Beijing 100190, China}

\author{Runxin Li}
\affiliation{State Key Laboratory of Quantum Functional Materials, Department of Physics,\\Guangdong Provincial Key Laboratory of Topological Matter, \\ Guangdong Basic Research Center of Excellence for Quantum Science and, \\ Center for Advanced Light Source, Southern University of Science and Technology (SUSTech), Shenzhen 518055, China}

\author{Yuchen Huo}
\affiliation{Department of Physics, Tsinghua University, Beijing 100084, China}
\affiliation{State Key Laboratory of Quantum Functional Materials, Department of Physics,\\Guangdong Provincial Key Laboratory of Topological Matter, \\ Guangdong Basic Research Center of Excellence for Quantum Science and, \\ Center for Advanced Light Source, Southern University of Science and Technology (SUSTech), Shenzhen 518055, China}

\author{Miaomiao Wang}
\affiliation{State Key Laboratory of Quantum Functional Materials, Department of Physics,\\Guangdong Provincial Key Laboratory of Topological Matter, \\ Guangdong Basic Research Center of Excellence for Quantum Science and, \\ Center for Advanced Light Source, Southern University of Science and Technology (SUSTech), Shenzhen 518055, China}

\author{Quanxing Wei}
\affiliation{State Key Laboratory of Quantum Functional Materials, Department of Physics,\\Guangdong Provincial Key Laboratory of Topological Matter, \\ Guangdong Basic Research Center of Excellence for Quantum Science and, \\ Center for Advanced Light Source, Southern University of Science and Technology (SUSTech), Shenzhen 518055, China}
 \affiliation{Quantum Science Center of Guangdong-Hong Kong-Macao Greater Bay Area, Shenzhen 518055, China}

 \author{Guihao Jia}
 \affiliation{State Key Laboratory of Quantum Functional Materials, Department of Physics,\\Guangdong Provincial Key Laboratory of Topological Matter, \\ Guangdong Basic Research Center of Excellence for Quantum Science and, \\ Center for Advanced Light Source, Southern University of Science and Technology (SUSTech), Shenzhen 518055, China}

\author{Qing Qu}
\affiliation{State Key Laboratory of Quantum Functional Materials, Department of Physics,\\Guangdong Provincial Key Laboratory of Topological Matter, \\ Guangdong Basic Research Center of Excellence for Quantum Science and, \\ Center for Advanced Light Source, Southern University of Science and Technology (SUSTech), Shenzhen 518055, China}

\author{Cui Ding}
\affiliation{Quantum Science Center of Guangdong-Hong Kong-Macao Greater Bay Area, Shenzhen 518055, China}

\author{Lin Zhao}
\affiliation{Beijing National Laboratory for Condensed Matter Physics and,\\ Institute of Physics, Chinese Academy of Sciences, Beijing 100190, China}
\affiliation{School of Physical Sciences, University of Chinese Academy of Sciences, Beijing 100190, China}

\author{Zhuoyu Chen}
\affiliation{State Key Laboratory of Quantum Functional Materials, Department of Physics,\\Guangdong Provincial Key Laboratory of Topological Matter, \\ Guangdong Basic Research Center of Excellence for Quantum Science and, \\ Center for Advanced Light Source, Southern University of Science and Technology (SUSTech), Shenzhen 518055, China}
\affiliation{Quantum Science Center of Guangdong-Hong Kong-Macao Greater Bay Area, Shenzhen 518055, China}

\author{Ming-Qiang Ren}
\email{renmq@sustech.edu.cn}
\affiliation{State Key Laboratory of Quantum Functional Materials, Department of Physics,\\Guangdong Provincial Key Laboratory of Topological Matter, \\ Guangdong Basic Research Center of Excellence for Quantum Science and, \\ Center for Advanced Light Source, Southern University of Science and Technology (SUSTech), Shenzhen 518055, China}
    
 \author{Kun Jiang}
 \email{jiangkun@iphy.ac.cn}
 \affiliation{Beijing National Laboratory for Condensed Matter Physics and,\\ Institute of Physics, Chinese Academy of Sciences, Beijing 100190, China}
 \affiliation{School of Physical Sciences, University of Chinese Academy of Sciences, Beijing 100190, China}

\author{X. J. Zhou}
\affiliation{Beijing National Laboratory for Condensed Matter Physics and,\\ Institute of Physics, Chinese Academy of Sciences, Beijing 100190, China}
\affiliation{School of Physical Sciences, University of Chinese Academy of Sciences, Beijing 100190, China}

\author{Jinfeng Jia}
\affiliation{State Key Laboratory of Quantum Functional Materials, Department of Physics,\\Guangdong Provincial Key Laboratory of Topological Matter, \\ Guangdong Basic Research Center of Excellence for Quantum Science and, \\ Center for Advanced Light Source, Southern University of Science and Technology (SUSTech), Shenzhen 518055, China}
\affiliation{Quantum Science Center of Guangdong-Hong Kong-Macao Greater Bay Area, Shenzhen 518055, China}

 \author{Qi-Kun Xue}
\affiliation{State Key Laboratory of Quantum Functional Materials, Department of Physics,\\Guangdong Provincial Key Laboratory of Topological Matter, \\ Guangdong Basic Research Center of Excellence for Quantum Science and, \\ Center for Advanced Light Source, Southern University of Science and Technology (SUSTech), Shenzhen 518055, China}
 \affiliation{Quantum Science Center of Guangdong-Hong Kong-Macao Greater Bay Area, Shenzhen 518055, China}
 \affiliation{Frontier Science Center for Quantum Information, Beijing 100084, China}
 \affiliation{Beijing Academy of Quantum Information Sciences, Beijing 100193, China}

\author{Yu-Jie Sun}
\email{sunyj@sustech.edu.cn}
\affiliation{State Key Laboratory of Quantum Functional Materials, Department of Physics,\\Guangdong Provincial Key Laboratory of Topological Matter, \\ Guangdong Basic Research Center of Excellence for Quantum Science and, \\ Center for Advanced Light Source, Southern University of Science and Technology (SUSTech), Shenzhen 518055, China}
\affiliation{Quantum Science Center of Guangdong-Hong Kong-Macao Greater Bay Area, Shenzhen 518055, China}

\date{\today}

\begin{abstract}
    The predominantly $d_{x^2-y^2}$-wave pairing symmetry of cuprate superconductors is well established, yet its microscopic manifestation in local Cooper-pair tunneling remains less explored. Here, we investigate Bi$_{1.2}$Pb$_{0.8}$Sr$_2$CaCu$_2$O$_{8+\delta}$, using Josephson scanning tunneling spectroscopy with a conventional $s$-wave superconducting Pb tip. We observe a zero-bias conductance enhancement whose approximately quadratic dependence on junction conductance is consistent with phase-diffusive Josephson tunneling. Atomic-resolution conductance maps and site-resolved spectra reveal a pronounced intra-unit-cell modulation, with enhanced Josephson response near oxygen sites and strong suppression near copper sites and along diagonal directions. Microscopic calculations incorporating the bilayer electronic structure and realistic Wannier-based tunneling matrix elements reproduce the observed spatial pattern through interference between pairing contributions of opposite signs. Our results reveal how the sign-changing $d$-wave superconducting order parameter manifests itself in atomic-scale Josephson tunneling, providing complementary evidence for unconventional pairing symmetry in cuprates.
\end{abstract}

\maketitle

%\textit{Introduction}

Since the discovery of high-temperature superconductivity, cuprates have played a central role in the study of strongly correlated electron systems and unconventional superconductivity \cite{Bednorz1986,LeeNagaosaWen2006,PWAnderson_2004,Keimer2015}. One of the most important established properties of cuprates is their predominantly $d_{x^2-y^2}$-wave superconducting pairing symmetry, characterized by a sign change of the order parameter between the two orthogonal Cu--O bond directions \cite{Scalapino1995,Tsuei2000,Xiang2022,Scalapino1986,Bickers1987,Gros1987,LeeRead1987,Kotliar1988, KotliarLiu1988}. The most direct evidence comes from phase-sensitive Josephson experiments. Corner-junction measurements revealed a $\pi$ phase shift between orthogonal crystallographic directions, while tricrystal experiments demonstrated spontaneous half-integer flux quantization, providing compelling evidence for a sign-changing superconducting order parameter \cite{Wollman1993,Tsuei1994,Tsuei2000}. More recently, Josephson experiments on twisted cuprates junctions have reported contrasting behavior near a $45^\circ$ twist angle, ranging from strongly suppressed Josephson coupling consistent with $d$-wave pairing to substantial coupling suggesting additional pairing components \cite{BSCCO_PRX,jongyun2021twisted,martini2023twisted,ZhaoTRSB2023}.
Complementary evidence has been obtained from various spectroscopic and thermodynamic measurements. The linear temperature dependence of the magnetic penetration depth at low temperatures provides evidence for nodal quasiparticle excitations \cite{Hardy1993}. Angle-resolved photoemission spectroscopy (ARPES) revealed a strongly anisotropic superconducting gap consistent with the $d_{x^2-y^2}$-wave form \cite{Shen1993,Ding1996,APRES_RevModPhys.75.473}. 
Furthermore, scanning tunneling microscopy and spectroscopy (STM/STS) have revealed a characteristic V-shaped low-energy density of states and quasiparticle interference patterns, providing complementary evidence for the nodal gap structure of $d$-wave superconductivity in cuprates \cite{Hudson1999,Pan2000,Hoffman2002,McElroy2003,STM_RevModPhys.79.353}. Together, these experimental findings have firmly established predominantly $d_{x^2-y^2}$-wave pairing as a defining characteristic of cuprate superconductivity.

Although the predominantly $d_{x^2-y^2}$-wave pairing symmetry of cuprates is well established, different experimental techniques probe distinct aspects of the superconducting state. Josephson scanning tunneling microscopy (JSTM) combines the sensitivity of Josephson tunneling to superconducting pair correlations with the atomic-scale spatial resolution of scanning tunneling microscopy, providing a complementary approach to investigating superconducting pairing symmetry \cite{Pan1998,Smakov2001,Naaman2001}. In particular, the Josephson coupling depends on the relative phases of the superconducting order parameter and the microscopic tunneling matrix elements, allowing interference between different pairing components to produce characteristic spatial patterns. Recently, JSTM has been applied to monolayer FeSe/SrTiO$_3$, revealing its nodeless $d$-wave pairing state \cite{Li2026FeSe}. This study demonstrates the capability of JSTM to probe the microscopic phase structure of unconventional superconductors and motivates further investigations of the superconducting pairing state in cuprates.

%JSTM has been employed to investigate the superconducting condensate in cuprates. Experimentally, superconducting Pb tip and  Bi$_2$Sr$_2$CaCu$_2$O$_{8+\delta}$ (Bi2212) nano-flake tips have been applied to probe $d$-wave cuprates Bi2212 \cite{Kimura2008, Hamidian2016,Joo2019}.
%Theoretically, the microscopic link JSTM has been proposed by Šmakov et al. \cite{Smakov2001}, Graham and Morr\cite{Graham2019}, Choubey and Hirschfeld\cite{Choubey2024}. However, the JSTM probe cuprates pairing symmetry remains highly unexplored.

JSTM has been employed to investigate the superconducting condensate in cuprates. Experimentally, superconducting Pb tips and Bi$_2$Sr$_2$CaCu$_2$O$_{8+\delta}$ (Bi2212) nanoflake tips have been used to probe local Cooper-pair tunneling in $d$-wave cuprates \cite{Kimura2008,Hamidian2016,Joo2019}. Theoretically, Šmakov et al. first proposed JSTM as a probe of the spatial structure of superconducting order parameters \cite{Smakov2001}, while subsequent studies by Graham and Morr \cite{Graham2017,Graham2019} and Choubey and Hirschfeld \cite{Choubey2024} explored its sensitivity to unconventional pairing symmetry and the microscopic Josephson response in $d$-wave cuprates. Despite these advances, the application of JSTM to directly investigate the microscopic signatures of $d$-wave pairing symmetry in cuprates remains relatively unexplored.

In this work, we investigate the superconducting pairing state of Bi$_{1.2}$Pb$_{0.8}$Sr$_2$CaCu$_2$O$_{8+\delta}$ (Pb-doped Bi2212), using Josephson scanning tunneling spectroscopy. Pb substitution strongly suppresses the incommensurate structural supermodulation of pristine Bi2212, providing a suitable platform for investigating its superconducting properties \cite{Fournier1996,Chong1997,Kendziora1997,PbSubstitution2019}. By combining conventional quasiparticle tunneling spectroscopy with Josephson tunneling measurements, we characterize both the superconducting excitation gap and the Cooper-pair tunneling response. We examine the Josephson coupling in the context of the established $d_{x^2-y^2}$-wave pairing symmetry and discuss its relationship with the unconventional superconducting order parameter. Our results provide complementary experimental insight into the superconducting pairing state of cuprates through local Cooper-pair tunneling measurements.

Single crystals of Pb-doped Bi2212 were cleaved \textit{in situ} under ultrahigh vacuum at room temperature and subsequently transferred to the scanning tunneling microscope head for measurements at 1.6 K under a pressure below $8\times10^{-11}$ Torr. Superconducting Pb tips were prepared by controlled indentation of PtIr tips into Pb films grown on Si(111) substrates. The superconducting character of the tips was verified by tunneling spectroscopy on the Pb films. Topographic images were acquired in constant-current mode, and differential conductance spectra and maps were recorded using lock-in detection at 973 Hz. For spatially resolved Josephson measurements, the junction was stabilized at the specified set point before disabling the feedback, and the zero-bias differential conductance, $g_0(\boldsymbol{r})$, was recorded as the local Josephson response.

\begin{figure}%[htbp!]
	\begin{center}
		\includegraphics[width=\columnwidth]{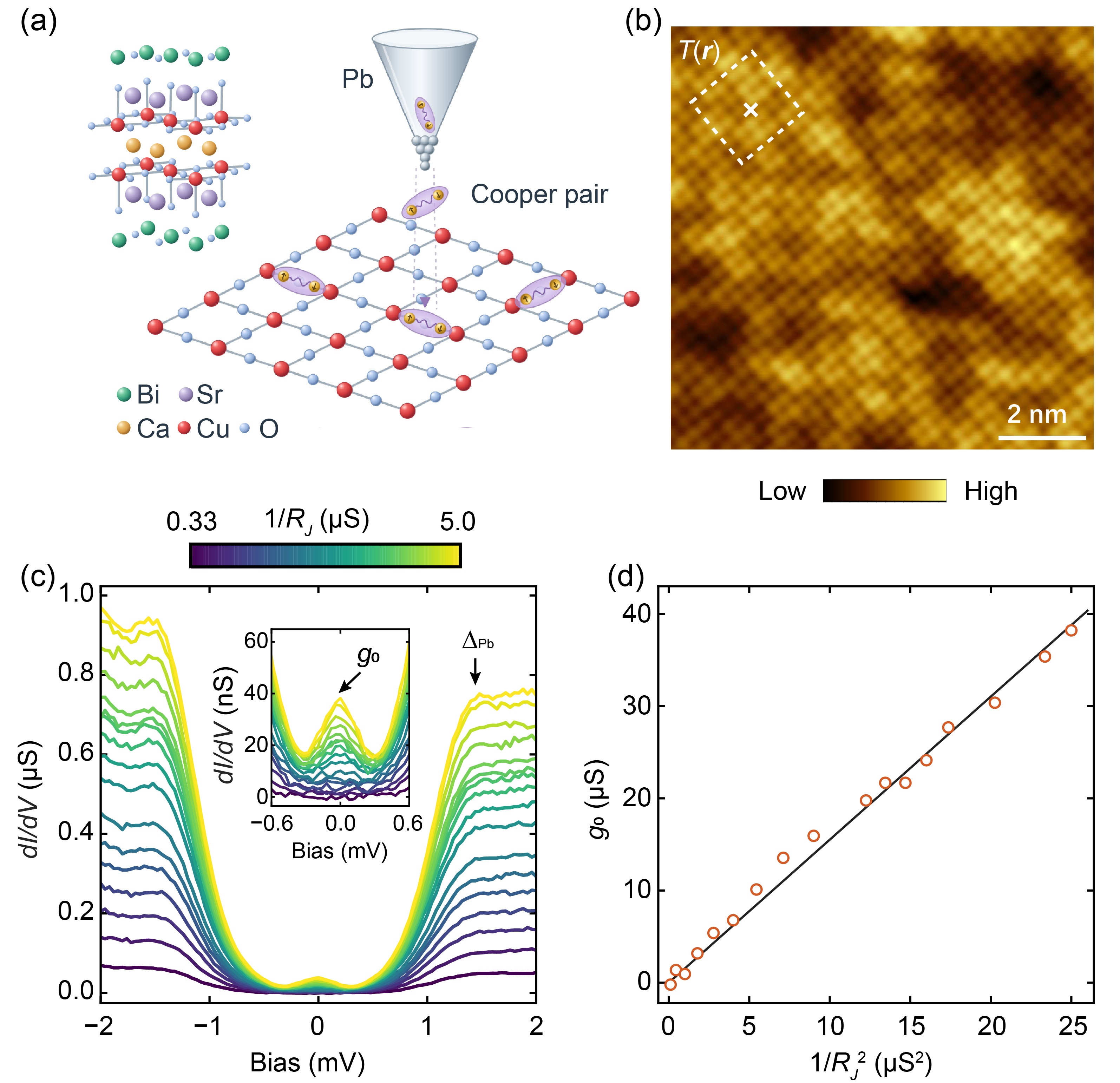}
		\caption{\textbf{Josephson tunneling with a superconducting Pb tip on Pb-doped Bi2212.}
        (a) Layered crystal structure of Bi2212 and schematic of a superconducting Pb tip probing the copper--oxygen plane. Red and light-blue spheres represent Cu and O, respectively; gold electron pairs illustrate pair tunneling schematically.
        (b) Atomically resolved topography acquired at a setpoint bias of 30~mV and a setpoint current of 200~pA. The dashed box and white cross indicate the selected region and spectroscopy position, respectively.
        (c) Low-energy tunneling spectra recorded at the position marked by the white cross in (b), with a fixed setpoint bias of 30~mV and setpoint currents of 10--150~nA. Color denotes the inverse junction resistance, $1/R_J$, with the junction resistance $R_J$ defined as the absolute ratio of setpoint bias to setpoint current and ranging from 0.20 to 3.00~megohms. The arrow indicates the superconducting gap edge of the Pb-functionalized tip, $\Delta_{\mathrm{Pb}}\approx 1.4$~meV. The inset magnifies the near-zero-bias region, showing an Josephson  conductance enhancement ($g_0$),  that grows with increasing junction conductance,$1/R_J$.
        (d) Zero-bias conductance extracted from (c) and plotted against the inverse square of the junction resistance,$1/R_J^2$. Open circles denote the measurements, and the solid line is a linear fit constrained through the origin. The approximately quadratic dependence on junction conductance is consistent with a phase-diffusive Josephson response. Measurements in (c,d) were recorded with a modulation amplitude of 200~uV.}
    \label{fig1}
	\end{center}
	%	\vskip-0.5cm
\end{figure}

%\textit{JSTM}
We first characterize the Josephson tunneling response of Pb-doped Bi2212. As illustrated in Fig.~\ref{fig1}(a), a conventional $s$-wave superconducting Pb tip with a gap of 1.4~meV forms a vacuum Josephson junction with the cuprate sample, enabling spatially resolved measurements of superconducting pair correlations. In Bi2212, the superconducting CuO$_2$ bilayers lie beneath the cleaved Bi(Pb)O surface, whose atomically resolved topography is shown in Fig.~\ref{fig1}(b). The atomic lattice is clearly resolved against a slowly varying topographic background, providing a crystallographic reference for spatially resolved tunneling measurements. Compared with pristine Bi2212 \cite{Hudson1999,Pan2000}, Pb substitution strongly suppresses the characteristic incommensurate structural supermodulation \cite{PbSubstitution2019}, facilitating the investigation of the local Josephson response at different positions within the unit cell.

We next examine the tunneling spectra acquired at the representative position marked by the white cross in Fig.~\ref{fig1}(b). Figure~\ref{fig1}(c) presents low-energy spectra measured with the superconducting Pb tip at 1.6~K, using a fixed setpoint bias of 30~mV and setpoint currents of 10--150~nA. 
%The plotted conductance is the recorded lock-in X current divided by the nominal modulation amplitude, and the junction resistance is estimated from the setpoint:
%$$
%G_X(V)=\frac{I_X(V)}{V_{\mathrm{mod}}},
%\qquad
%R_J=\left|\frac{V_{\mathrm{set}}}{I_{\mathrm{set}}}\right|.
%$$

\begin{figure}%[htbp!]
	\begin{center}
		\includegraphics[width=\columnwidth]{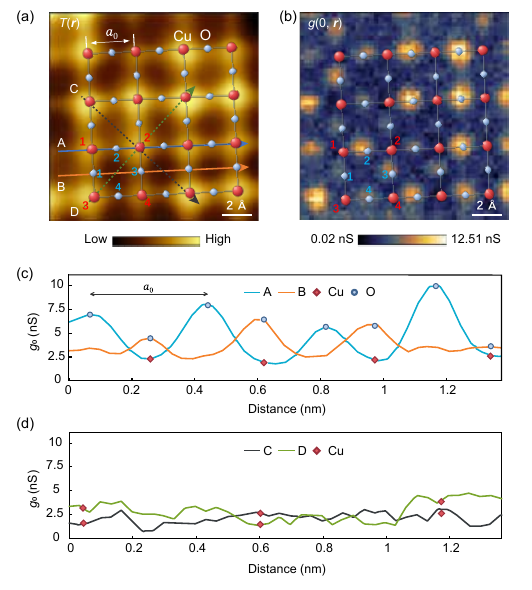}
		\caption{\textbf{Intra-unit-cell variation of the zero-bias Josephson  response in Pb-doped Bi2212.}
        (a,b) Atomically resolved topography (a) and zero-bias Josephson conductance map (b) of the region outlined by the white dashed box in Fig.~\ref{fig1}(b), acquired at a setpoint bias of 30~mV and a setpoint current of 85~nA. The assigned Cu (red) and O (light-blue) positions are overlaid. Blue, orange, black, and green arrows indicate paths A, B, C, and D, respectively, with A and B along the parallel directions and C and D along the diagonal directions. The numbered sites indicate the positions used for the spectra in Fig.~\ref{fig3}(a,b).The double-headed arrows in (a) indicate the nearest-neighbor Cu--Cu spacing. 
        (c) Zero-bias conductance profiles along A and B. Light-blue circles and filled red diamonds mark O and Cu sites, respectively. Conductance maxima occur near the marked O sites, whereas the response is substantially lower near Cu sites. The double-headed arrow indicates the nearest-neighbor Cu--Cu spacing.
        (d) Corresponding profiles along C and D, both showing a weak but finite zero-bias conductance attributed to quasiparticle background. Open red diamonds mark schematic Cu-site positions along the profiles.}
    \label{fig2}
	\end{center}
	%	\vskip-0.5cm
\end{figure}

\begin{figure}%[htbp!]
    \begin{center}
        \includegraphics[width=\columnwidth]{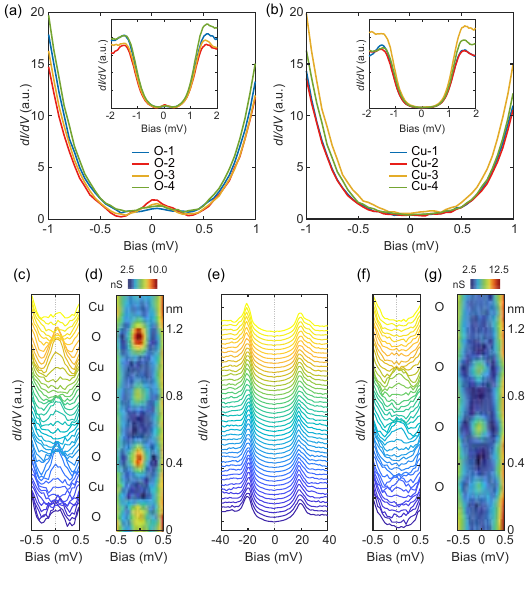}
        \caption{\textbf{Site-resolved and spatially resolved low-energy tunneling spectra in Pb-doped Bi2212.}
        (a,b) Representative spectra at the four numbered O sites (a) and four numbered Cu sites (b) marked in Fig.~\ref{fig2}(a,b), acquired at a setpoint bias of 30~mV and a setpoint current of 85~nA. The Josephson enhancement is more pronounced at the selected O sites and much weaker at Cu sites.Insets show the corresponding spectra over a wider bias range of $\pm 2$~mV.
        (c--e) Spatially resolved spectra along path A indicated by the blue arrow in Fig.~\ref{fig2}(a), shown as a near-zero-bias spectral stack (c), a distance--bias conductance map (d), and a wide-bias spectral stack (e).
        (f,g) Corresponding near-zero-bias spectral stack (f) and distance--bias map (g) along path B, indicated by the orange arrow in Fig.~\ref{fig2}(a). The Cu and O labels mark the corresponding atomic positions along the scan paths. Successive spectra in (c,e,f) are offset vertically for clarity.}
        \label{fig3}
    \end{center}
\end{figure}

As the junction conductance, $1/R_J$, increases, a pronounced zero-bias enhancement emerges and grows in intensity, as highlighted in the inset of Fig.~\ref{fig1}(c). This feature is consistent with Josephson tunneling, arising from the transfer of Cooper pairs between the superconducting tip and sample. In STM junctions, the Josephson coupling energy can be smaller than the thermal energy; phase fluctuations then give rise to a broadened low-bias Josephson response rather than a dissipationless dc supercurrent \cite{Ivanchenko1969,Naaman2001}.

To further examine the Josephson origin of the zero-bias enhancement, we investigate its dependence on the setpoint junction resistance. Within the Ivanchenko--Zil'berman model, the Josephson contribution to the zero-bias conductance, $g_0$, in the phase-diffusive regime is proportional to the square of the critical current \cite{Ivanchenko1969,Naaman2001}. For a fixed electromagnetic environment and tunneling geometry, the expected junction-resistance dependence is $g_0\propto I_c^2\propto\frac{1}{R_J^2}.$ As shown in Fig.~\ref{fig1}(d), the measured zero-bias conductance exhibits an approximately linear dependence on the inverse square of the junction resistance; the solid line is a linear fit constrained through the origin. This behavior is consistent with phase-diffusive Josephson tunneling and provides further support for the Cooper-pair tunneling origin of the zero-bias enhancement.

%These measurements demonstrate local Josephson tunneling between a conventional $s$-wave Pb tip and the $d$-wave cuprate Bi2212. Although the Josephson coupling between ideal $s$-wave and $d$-wave superconductors can vanish due to symmetry cancellation, the spatially localized tunneling geometry of STM allows a finite coupling through microscopic tunneling processes \cite{Choubey2024}. The observed Josephson response thus provides an experimental foundation for further investigating the unconventional superconducting pairing state of cuprates.

%\textit{Intra-unit-cell Josephson response}
Having established the Josephson tunneling response in Pb-doped Bi2212, we next investigate its spatial variation within the crystallographic unit cell. 
Figures~\ref{fig2}(a) and \ref{fig2}(b) show the atomically resolved topography $T(\boldsymbol{r})$ and the corresponding zero-bias conductance map,$g(0,\boldsymbol{r})$, respectively, acquired in the region highlighted in Fig.~\ref{fig1}(b). 
The assigned Cu and O positions in the underlying CuO$_2$ plane are superimposed on both images to establish the correspondence between the local tunneling response and the crystallographic lattice. Remarkably, the zero-bias conductance exhibits a pronounced intra-unit-cell modulation, with enhanced intensity near the O sites and a substantially suppressed response near the Cu sites.

To quantify this spatial modulation, we examine the zero-bias conductance profiles along two parallel paths, A and B, indicated in Figs.~\ref{fig2}(a) and \ref{fig2}(b). As shown in Fig.~\ref{fig2}(c), both profiles exhibit pronounced variations on the scale of the nearest-neighbor Cu--Cu distance $a_0$. The conductance maxima are predominantly located near the O sites, whereas the response is suppressed near the Cu sites. This spatial correspondence is observed over multiple unit cells, demonstrating a systematic dependence of the local zero-bias tunneling response on the atomic positions within the CuO$_2$ lattice.

We further examine the spatial dependence along the diagonal paths C and D, as illustrated in Fig.~\ref{fig2}(a). In contrast to the pronounced modulation along the Cu--O bond directions, the corresponding conductance profiles in Fig.~\ref{fig2}(d) exhibit a much weaker response, without the prominent maxima observed near the O sites along paths A and B. The conductance along these diagonal directions remains finite, which may be attributed to a residual quasiparticle tunneling background. The contrast between the bond-direction and diagonal profiles reveals a strong directional dependence of the intra-unit-cell zero-bias conductance.

%These observations demonstrate that the local tunneling response is strongly influenced by the microscopic lattice positions, with enhanced conductance near O sites and suppressed conductance near Cu sites and along the diagonal directions. Such a spatial structure is particularly interesting for the $d_{x^2-y^2}$-wave superconducting state of cuprates, where the pairing amplitudes along the two orthogonal Cu--O bond directions have opposite signs. Microscopic calculations have predicted that interference between these pairing components can produce a strongly site-dependent Josephson coupling \cite{Choubey2024}. To further clarify the origin of the observed intra-unit-cell modulation and distinguish the Josephson contribution from the quasiparticle background, we next examine the tunneling spectra at representative Cu and O positions.

To further investigate the microscopic origin of the intra-unit-cell modulation of the zero-bias conductance, we examine the local tunneling spectra at representative O and Cu sites. Figures~\ref{fig3}(a) and \ref{fig3}(b) present the differential-conductance spectra acquired at the O$_1$--O$_4$ and Cu$_1$--Cu$_4$ positions marked in Fig.~\ref{fig2}(a), respectively. Although both sets of spectra exhibit characteristic superconducting quasiparticle features, their low-bias responses show a striking difference. As highlighted in the insets, a pronounced zero-bias conductance peak is observed at the selected O sites, whereas the corresponding feature is strongly suppressed at the Cu sites. This contrast is consistent with the oxygen-centered enhancement of the zero-bias conductance observed in Fig.~\ref{fig2}.

To examine the spatial evolution of the tunneling spectra, we perform a continuous line scan along path A in Fig.~\ref{fig2}(a). Figures~\ref{fig3}(c)--\ref{fig3}(e) present the near-zero-bias spectral stack, the corresponding distance--bias conductance map, and the wide-bias spectral stack, respectively. The zero-bias Josephson peak exhibits pronounced spatial variations on the scale of the Cu--Cu
lattice spacing $a_0$. This modulation is particularly evident in Fig.~\ref{fig3}(d), where enhanced conductance is concentrated near zero bias and periodically distributed along the scanning direction. The conductance maxima coincide with the O sites, consistent with the spatial distribution of the Josephson response in Fig.~\ref{fig2}(c). The wide-bias spectra in Fig.~\ref{fig3}(e) exhibit pronounced superconducting coherence peaks at approximately $\pm 20$~meV of Pb-doped Bi2212. The well-defined coherence peaks indicate that the measurements probe a superconducting region rather than one dominated by the pseudogap.

Similar behavior is observed along the parallel path B. As shown in Figs~\ref{fig3}(f) and \ref{fig3}(g), the near-zero-bias spectral and corresponding distance--bias map  reveal a spatially modulated zero-bias response, with enhanced conductance near selected O positions. Although the detailed peak amplitudes vary between the two paths, the recurring site-dependent contrast demonstrates that the intra-unit-cell modulation is reproducible across multiple lattice periods.

Together, Figs.~\ref{fig2} and \ref{fig3} establish a pronounced intra-unit-cell variation of the low-energy tunneling response in Pb-doped Bi2212. In particular, the oxygen-centered enhancement and copper-site suppression are associated with the spatial evolution of the narrow zero-bias peak, consistent with a strongly site-dependent Josephson contribution. Such behavior is particularly interesting for $d_{x^2-y^2}$-wave superconductivity, where the sign-changing pairing amplitudes can give rise to constructive or destructive interference in local Cooper-pair tunneling \cite{Choubey2024}. These observations motivate a microscopic interpretation of the spatial structure of the Josephson response in cuprates.

%\textit{Theory}
To interpret the intra-unit-cell modulation of the Josephson response, we consider an $s$-wave superconducting tip coupled to the CuO$_2$ bilayer of Bi2212. The sample is described by a bilayer tight-binding Hamiltonian with parameters reported for Bi2212~\cite{Drozdov2018}. We introduce intralayer pairing with the same condensate phase in both layers and the gap function $\Delta(\bk)=\Delta_0[\cos(k_xa)-\cos(k_ya)]/2$, where the $x$ and $y$ axes lie along the Cu--O bonds and $a$ is the nearest-neighbor Cu--Cu spacing. The spatial tunneling amplitudes are constructed from the Cu $d_{x^2-y^2}$ Wannier functions $w_i(\br)$ of Choubey \textit{et al.}~\cite{Choubey2017}, with the tip coupled coherently to both layers.

In the weak-tunneling limit, the local critical Josephson current $I_c(\br)=(2e/\hbar)|\mathcal J(\br)|$ is determined by the coherent
pair-transfer amplitude~\cite{Choubey2024},
\begin{equation}
	\mathcal J(\br)=k_BTt_0^2\sum_{n,ij}w_i(\br)w_j(\br)
	\operatorname{tr}_{\mathrm{spin}}\!\left[
	\bar F_t^0(i\omega_n)F_{s,ij}^0(i\omega_n)\right].
	\label{eq:theory-josephson}
\end{equation}
Here $i,j$ run over Cu sites in both layers, and $t_0w_i(\br)$ is the real, spin-independent tunneling amplitude between the tip and site $i$; the prefactor $t_0$ is common to both layers. The fermionic Matsubara frequencies are $\omega_n=(2n+1)\pi k_BT$. The sample and tip anomalous Green functions, $F_{s,ij}^0$ and $\bar F_t^0$, are defined with their condensate phases removed, and $\operatorname{tr}_{\mathrm{spin}}$ denotes the trace over spin indices. Contributions from different sites and layers are summed before taking the magnitude, allowing pair-tunneling amplitudes of opposite sign to cancel.

\begin{figure}[t]
	\centering
	\includegraphics[width=\columnwidth]{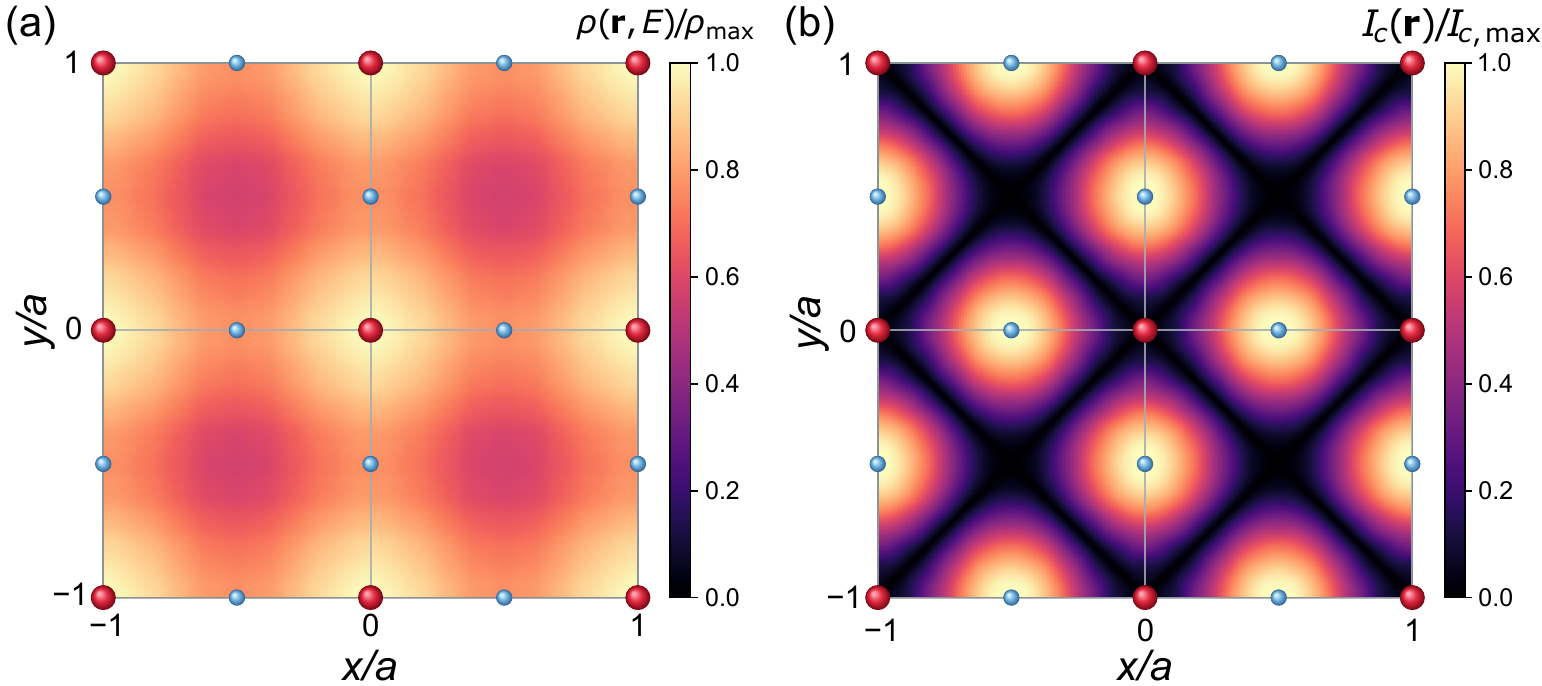}
	\caption{\textbf{Calculated intra-unit-cell tunneling response of bilayer Bi2212.}
        (a) Spin-summed local density of states (LDOS) at zero energy and
        (b) Josephson critical-current magnitude for the bilayer $d$-wave
        state. Each map is normalized to its own maximum. Red and light-blue
        spheres mark the in-plane Cu and O positions, respectively, with
        $a=3.83$~\AA. The tip is positioned approximately $5$~\AA\ above the
        BiO surface. The calculation uses sample and tip gap amplitudes of
        $\Delta_0=19$~meV and $\Delta_t=1.4$~meV, respectively, and
        $T=1.6$~K.}
	\label{fig:theory-jstm}
\end{figure}

Figure~\ref{fig:theory-jstm} compares the calculated zero-energy LDOS
with the Josephson critical-current magnitude. The LDOS is largest above
Cu sites, where the Josephson response is strongly suppressed, while the
critical current reaches its maxima near the O positions.
These different spatial patterns reflect the distinct correlations probed
by the two observables: the LDOS is determined by the normal Green function,
whereas the Josephson response depends on the coherent sum of anomalous
pair amplitudes in Eq.~\eqref{eq:theory-josephson}.

The suppression along the diagonals follows from the symmetry of the
pairing state and the tunneling geometry. For an ideal
reflection-symmetric junction, exchanging $x$ and $y$ maps the products
$w_iw_j$ onto those of symmetry-related sites, while reversing the sign
of the $d$-wave anomalous amplitudes. The pair-transfer amplitude therefore
satisfies $\mathcal J(y,x)=-\mathcal J(x,y)$, enforcing cancellation along $x=y$. Symmetry-related diagonals exhibit the
same cancellation, accounting for the diagonal minima and Cu-site
suppression in Fig.~\ref{fig:theory-jstm}(b). As the tip moves toward an
O position, the opposite-sign contributions acquire unequal tunneling
weights, lifting the cancellation and producing a finite Josephson
coupling. The calculated pattern reproduces the principal spatial
features of the measurements, providing a microscopic interpretation of
the bond-centered enhancement and diagonal suppression in terms of the
$d_{x^2-y^2}$-wave pairing state of Bi2212.

In summary, we have investigated the local Josephson tunneling response of Pb-doped Bi2212 using a superconducting Pb tip. The zero-bias conductance peak and its approximately quadratic dependence on junction conductance are consistent with phase-diffusive Josephson tunneling. Atomic-resolution measurements reveal a pronounced intra-unit-cell modulation, with enhanced Josephson response near O sites and strong suppression near Cu sites and along diagonal directions. Microscopic calculations incorporating the bilayer electronic structure and realistic Wannier functions reproduce the observed spatial pattern through interference between the opposite-sign pairing components of the $d_{x^2-y^2}$-wave superconducting order parameter. The position-dependent tunneling amplitudes control the interference between these components, giving rise to the characteristic spatial modulation of the Josephson response. Our results provide microscopic evidence consistent with the sign-changing $d$-wave pairing state of cuprates and establish a complementary connection between superconducting pairing symmetry and local Cooper-pair tunneling.

{\em Note added.—} During the final preparation of this manuscript, we became aware of an independent study reporting similar observations of intra-unit-cell Josephson tunneling in copper and iron based superconductors \cite{wei2026directrealspaceimagingsignreversal}.

{\em Acknowledgments.}-- We acknowledge the support by the National Natural Science Foundation of China (12304164, 12674164), the Ministry of Science and Technology (Grant 2024YFA1408102, 2022YFA1403101), the Chinese Academy of Sciences Project for Young Scientists in Basic Research (2022YSBR-048), the New Cornerstone Investigator Program,  the Guangdong Provincial Key Laboratory of Topological Matter (2026B1212010004), the Shenzhen Natural Science Foundation (JCYJ20240813094203005).

\bibliographystyle{apsrev4-2}
\bibliography{ref}

@article{Chong1997,
  author = {Chong, I. and Hiroi, Z. and Izumi, M. and
            Shimoyama, J. and Nakayama, Y. and Kishio, K. and
            Terashima, T. and Bando, Y. and Takano, M.},
  title = {High Critical-Current Density in the Heavily {Pb}-Doped
           {Bi$_2$Sr$_2$CaCu$_2$O$_{8+\delta}$} Superconductor:
           Generation of Efficient Pinning Centers},
  journal = {Science},
  volume = {276},
  number = {5313},
  pages = {770--773},
  year = {1997},
  doi = {10.1126/science.276.5313.770}
}

@article{Fournier1996,
  author = {Fournier, P. and Kapitulnik, A. and Marshall, A. F.},
  title = {Growth and annealing of
           {(Bi$_{1-x}$Pb$_x$)$_2$Sr$_2$CaCu$_2$O$_{8+\delta}$}
           single crystals},
  journal = {Physica C},
  volume = {257},
  number = {3-4},
  pages = {291--298},
  year = {1996},
  doi = {10.1016/0921-4534(95)00671-0}
}

@article{Kendziora1997,
  author = {Kendziora, C. and Qadri, S. B. and Skelton, E.},
  title = {{Pb} substitution in the
           {Bi$_2$Sr$_2$CaCu$_2$O$_{8+\delta}$} high-{$T_c$}
           superconductor: Cation overdoping},
  journal = {Phys. Rev. B},
  volume = {56},
  number = {22},
  pages = {14717--14722},
  year = {1997},
  doi = {10.1103/PhysRevB.56.14717}
}

@misc{wei2026directrealspaceimagingsignreversal,
      title={Direct real-space imaging of sign-reversal superconducting state by scanning Josephson tunneling}, 
      author={Tianheng Wei and Yanzhao Liu and Wei Ren and Zhen Liang and Ziqiang Wang and Jian Wang},
      year={2026},
      eprint={2610.38421},
      archivePrefix={arXiv},
      primaryClass={cond-mat.supr-con},
      url={https://arxiv.org/abs/2610.38421}, 
}

@article{BSCCO_PRX,
	title = {Presence of $s$-Wave Pairing in {Josephson} Junctions Made of Twisted Ultrathin {${\mathrm{Bi}}_{2}{\mathrm{Sr}}_{2}{\mathrm{CaCu}}_{2}{\mathrm{O}}_{8+x}$} Flakes},
	author = {Zhu, Yuying and Liao, Menghan and Zhang, Qinghua and Xie, Hong-Yi and Meng, Fanqi and Liu, Yaowu and Bai, Zhonghua and Ji, Shuaihua and Zhang, Jin and Jiang, Kaili and Zhong, Ruidan and Schneeloch, John and Gu, Genda and Gu, Lin and Ma, Xucun and Zhang, Ding and Xue, Qi-Kun},
	journal = {Phys. Rev. X},
	volume = {11},
	issue = {3},
	pages = {031011},
	numpages = {13},
	year = {2021},
	month = {Jul}, 
    doi = {10.1103/PhysRevX.11.031011},
    url = {https://link.aps.org/doi/10.1103/PhysRevX.11.031011},
	publisher = {American Physical Society}
}

@article{jongyun2021twisted,
author = {Lee, Jongyun and Lee, Wonjun and Kim, Gi-Yeop and Choi, Yong-Bin and Park, Jinho and Jang, Seong and Gu, Genda and Choi, Si-Young and Cho, Gil Young and Lee, Gil-Ho and Lee, Hu-Jong},
title = {{Twisted van der Waals Josephson Junction Based on a High-$T_c$ Superconductor}},
journal = {Nano Lett.},
volume = {21},
number = {24},
pages = {10469-10477},
year = {2021},
doi = {10.1021/acs.nanolett.1c03906}
}

@article{martini2023twisted,
  title={{Twisted cuprate van der Waals heterostructures with controlled Josephson coupling}},
  author={Martini, Mickey and Lee, Yejin and Confalone, Tommaso and Shokri, Sanaz and Saggau, Christian N and Wolf, Daniel and Gu, Genda and Watanabe, Kenji and Taniguchi, Takashi and Montemurro, Domenico and others},
  journal={Mater. Today},
  volume={67},
  pages={106--112},
  year={2023},
  publisher={Elsevier},
  doi = {https://doi.org/10.1016/j.mattod.2023.06.007},
  url = {https://www.sciencedirect.com/science/article/pii/S1369702123001980},
}

@article{ZhaoTRSB2023,
	title={Time-reversal symmetry breaking superconductivity between twisted cuprate superconductors},
	author={Zhao, SY Frank and Cui, Xiaomeng and Volkov, Pavel A and Yoo, Hyobin and Lee, Sangmin and Gardener, Jules A and Akey, Austin J and Engelke, Rebecca and Ronen, Yuval and Zhong, Ruidan and others},
	journal={Science},
	volume={382},
	number={6677},
	pages={1422--1427},
	year={2023},
    doi = {10.1126/science.abl8371},
URL = {https://www.science.org/doi/abs/10.1126/science.abl8371},
	publisher={American Association for the Advancement of Science}
}

@article{Ivanchenko1969,
	author = {Ivanchenko, Yu. M. and Zil'berman, L. A.},
	title = {The Josephson Effect in Small Tunnel Contacts},
	journal = {Sov. Phys. JETP},
	volume = {28},
	number = {6},
	pages = {1272--1276},
	year = {1969},
	url = {https://jetp.ras.ru/cgi-bin/dn/e_028_06_1272.pdf}
}

@article{LeeNagaosaWen2006,
  title = {Doping a Mott insulator: Physics of high-temperature superconductivity},
  author = {Lee, Patrick A. and Nagaosa, Naoto and Wen, Xiao-Gang},
  journal = {Rev. Mod. Phys.},
  volume = {78},
  issue = {1},
  pages = {17--85},
  numpages = {0},
  year = {2006},
  month = {Jan},
  publisher = {American Physical Society},
  doi = {10.1103/RevModPhys.78.17},
  url = {https://link.aps.org/doi/10.1103/RevModPhys.78.17}
}

@article{PWAnderson_2004,
doi = {10.1088/0953-8984/16/24/R02},
url = {https://doi.org/10.1088/0953-8984/16/24/R02},
year = {2004},
month = {jun},
publisher = {},
volume = {16},
number = {24},
pages = {R755},
author = {P W Anderson and P A Lee and M Randeria and T M Rice and N Trivedi and F C Zhang},
title = {The physics behind high-temperature superconducting cuprates: the ‘plain vanilla’ version
of RVB},
journal = {Journal of Physics: Condensed Matter}
}

@article{Scalapino1986,
  author = {Scalapino, D. J. and Loh, E., Jr.
            and Hirsch, J. E.},
  title = {{$d$}-Wave Pairing Near a
           Spin-Density-Wave Instability},
  journal = {Phys. Rev. B},
  volume = {34},
  pages = {8190},
  year = {1986},
  doi = {10.1103/PhysRevB.34.8190}
}

@article{Bickers1987,
  author = {Bickers, N. E. and Scalapino, D. J.
            and Scalettar, R. T.},
  title = {{CDW} and {SDW} Mediated
           Pairing Interactions},
  journal = {Int. J. Mod. Phys. B},
  volume = {1},
  pages = {687--695},
  year = {1987},
  doi = {10.1142/S0217979287001079}
}

@article{LeeRead1987,
  author = {Lee, P. A. and Read, N.},
  title = {Why Is {$T_c$} of the Oxide
           Superconductors So Low?},
  journal = {Phys. Rev. Lett.},
  volume = {58},
  pages = {2691},
  year = {1987},
  doi = {10.1103/PhysRevLett.58.2691}
}

@article{Kotliar1988,
  author = {Kotliar, G.},
  title = {Resonating Valence Bonds and
           {$d$}-Wave Superconductivity},
  journal = {Phys. Rev. B},
  volume = {37},
  pages = {3664},
  year = {1988},
  doi = {10.1103/PhysRevB.37.3664}
}

@article{KotliarLiu1988,
  author = {Kotliar, G. and Liu, J.},
  title = {Superexchange Mechanism and
           {$d$}-Wave Superconductivity},
  journal = {Phys. Rev. B},
  volume = {38},
  pages = {5142},
  year = {1988},
  doi = {10.1103/PhysRevB.38.5142}
}

@article{Gros1987,
	author = {Gros, C. and Joynt, R. and Rice, T. M.},
	title = {Superconducting Instability in the Large-{$U$}
	Limit of the Two-Dimensional Hubbard Model},
	journal = {Z. Phys. B},
	volume = {68},
	pages = {425--432},
	year = {1987},
	doi = {10.1007/BF01471072}
}

@book{Xiang2022,
  author = {Xiang, Tao and Wu, Congjun},
  title = {D-Wave Superconductivity},
  publisher = {Cambridge University Press},
  year = {2022},
  doi = {10.1017/9781009218566},
  isbn = {9781009218597}
}

@article{Pan1998,
  author = {Pan, S. H. and Hudson, E. W.
            and Davis, J. C.},
  title = {Vacuum Tunneling of Superconducting
           Quasiparticles from Atomically Sharp
           Scanning Tunneling Microscope Tips},
  journal = {Appl. Phys. Lett.},
  volume = {73},
  pages = {2992--2994},
  year = {1998},
  doi = {10.1063/1.122654}
}

@article{Naaman2001,
  author = {Naaman, O. and Teizer, W.
            and Dynes, R. C.},
  title = {Fluctuation Dominated Josephson
           Tunneling with a Scanning Tunneling
           Microscope},
  journal = {Phys. Rev. Lett.},
  volume = {87},
  pages = {097004},
  year = {2001},
  doi = {10.1103/PhysRevLett.87.097004}
}

@article{Smakov2001,
  author = {{\v S}makov, Jurij and Martin, Ivar
            and Balatsky, Alexander V.},
  title = {Josephson scanning tunneling microscopy},
  journal = {Phys. Rev. B},
  volume = {64},
  pages = {212506},
  year = {2001},
  doi = {10.1103/PhysRevB.64.212506}
}

@article{Choubey2024,
  author = {Choubey, Peayush and Hirschfeld, P. J.},
  title = {Theory of Josephson Scanning Microscopy with
           {$s$}-Wave Tip on Unconventional Superconducting
           Surface: Application to
           {Bi$_2$Sr$_2$CaCu$_2$O$_{8+\delta}$}},
  journal = {Phys. Rev. B},
  volume = {110},
  pages = {184519},
  year = {2024},
  doi = {10.1103/PhysRevB.110.184519}
}

@article{Graham2017,
  title = {Imaging the spatial form of a superconducting order parameter via Josephson scanning tunneling spectroscopy},
  author = {Graham, Martin and Morr, Dirk K.},
  journal = {Phys. Rev. B},
  volume = {96},
  issue = {18},
  pages = {184501},
  numpages = {7},
  year = {2017},
  month = {Nov},
  publisher = {American Physical Society},
  doi = {10.1103/PhysRevB.96.184501},
  url = {https://link.aps.org/doi/10.1103/PhysRevB.96.184501}
}

@article{Graham2019,
  title = {Josephson Scanning Tunneling Spectroscopy in ${d}_{{x}^{2}\ensuremath{-}{y}^{2}}$-Wave Superconductors: A Probe for the Nature of the Pseudogap in the Cuprate Superconductors},
  author = {Graham, Martin and Morr, Dirk K.},
  journal = {Phys. Rev. Lett.},
  volume = {123},
  issue = {1},
  pages = {017001},
  numpages = {6},
  year = {2019},
  month = {Jul},
  publisher = {American Physical Society},
  doi = {10.1103/PhysRevLett.123.017001},
  url = {https://link.aps.org/doi/10.1103/PhysRevLett.123.017001}
}

@article{APRES_RevModPhys.75.473,
  title = {Angle-resolved photoemission studies of the cuprate superconductors},
  author = {Damascelli, Andrea and Hussain, Zahid and Shen, Zhi-Xun},
  journal = {Rev. Mod. Phys.},
  volume = {75},
  issue = {2},
  pages = {473--541},
  numpages = {0},
  year = {2003},
  month = {Apr},
  publisher = {American Physical Society},
  doi = {10.1103/RevModPhys.75.473},
  url = {https://link.aps.org/doi/10.1103/RevModPhys.75.473}
}

@article{Hudson1999,
  author = {Hudson, E. W. and Pan, S. H. and
            Gupta, A. K. and Ng, K.-W. and Davis, J. C.},
  title = {Atomic-Scale Quasi-Particle Scattering
           Resonances in
           {Bi$_2$Sr$_2$CaCu$_2$O$_{8+\delta}$}},
  journal = {Science},
  volume = {285},
  number = {5424},
  pages = {88--91},
  year = {1999},
  doi = {10.1126/science.285.5424.88}
}

@article{Pan2000,
  author = {Pan, S. H. and Hudson, E. W. and
            Lang, K. M. and Eisaki, H. and
            Uchida, S. and Davis, J. C.},
  title = {Imaging the Effects of Individual Zinc
           Impurity Atoms on Superconductivity in
           {Bi$_2$Sr$_2$CaCu$_2$O$_{8+\delta}$}},
  journal = {Nature},
  volume = {403},
  pages = {746--750},
  year = {2000},
  doi = {10.1038/35001534}
}

@article{STM_RevModPhys.79.353,
  title = {Scanning tunneling spectroscopy of high-temperature superconductors},
  author = {Fischer, {\O{}}ystein and Kugler, Martin and Maggio-Aprile, Ivan and Berthod, Christophe and Renner, Christoph},
  journal = {Rev. Mod. Phys.},
  volume = {79},
  issue = {1},
  pages = {353--419},
  numpages = {0},
  year = {2007},
  month = {Mar},
  publisher = {American Physical Society},
  doi = {10.1103/RevModPhys.79.353},
  url = {https://link.aps.org/doi/10.1103/RevModPhys.79.353}
}

@misc{Li2026FeSe,
  author = {Li, Runxin and Huo, Yuchen and Mei, Jiong
            and Ren, Ming-Qiang and Wang, Miaomiao
            and Wei, Quanxing and Yao, Fengyu
            and Zheng, Maowen and Qu, Qing
            and Ding, Cui and Yuan, Hongtao
            and Chen, Zhuoyu and Chen, Wei-Qiang
            and Jiang, Kun and Xue, Qi-Kun
            and Sun, Yu-Jie},
  title = {Nodeless {$d$}-Wave Pairing in Monolayer
           {FeSe} Revealed by Intra-Unit-Cell
           Josephson Interferometry},
  year = {2026},
  eprint = {2610.34382},
  archivePrefix = {arXiv},
  primaryClass = {cond-mat.supr-con},
  url = {https://arxiv.org/abs/2610.34382}
}

@article{Bednorz1986,
  author = {Bednorz, J. G. and M{\"u}ller, K. A.},
  title = {Possible High {$T_c$} Superconductivity
           in the {Ba--La--Cu--O} System},
  journal = {Z. Phys. B},
  volume = {64},
  pages = {189--193},
  year = {1986},
  doi = {10.1007/BF01303701}
}

@article{Keimer2015,
  author = {Keimer, B. and Kivelson, S. A. and
            Norman, M. R. and Uchida, S. and Zaanen, J.},
  title = {From Quantum Matter to High-Temperature
           Superconductivity in Copper Oxides},
  journal = {Nature},
  volume = {518},
  pages = {179--186},
  year = {2015},
  doi = {10.1038/nature14165}
}

@article{Scalapino1995,
  author = {Scalapino, D. J.},
  title = {The Case for {$d_{x^2-y^2}$} Pairing
           in the Cuprate Superconductors},
  journal = {Phys. Rep.},
  volume = {250},
  pages = {329--365},
  year = {1995},
  doi = {10.1016/0370-1573(94)00086-I}
}

@article{Joo2019,
  author = {Joo, Sang Hyun and Kim, Jae-Joon and
            Yoo, Jung Hoon and Park, Min Seok and
            Lee, Kyung Seok and Gu, G. and Lee, Jinho},
  title = {Cooper Pair Density of
           {Bi$_2$Sr$_2$CaCu$_2$O$_{8+x}$}
           in Atomic Scale at 4.2 K},
  journal = {Nano Lett.},
  volume = {19},
  pages = {1112--1117},
  year = {2019},
  doi = {10.1021/acs.nanolett.8b04415}
}

@article{Hoffman2002,
  author = {Hoffman, J. E. and McElroy, K. and Lee, D.-H.
            and Lang, K. M. and Eisaki, H. and Uchida, S.
            and Davis, J. C.},
  title = {Imaging Quasiparticle Interference in
           {Bi$_2$Sr$_2$CaCu$_2$O$_{8+\delta}$}},
  journal = {Science},
  volume = {297},
  pages = {1148--1151},
  year = {2002},
  doi = {10.1126/science.1072640}
}

@article{McElroy2003,
  author = {McElroy, K. and Simmonds, R. W.
            and Hoffman, J. E. and Lee, D.-H.
            and Orenstein, J. and Eisaki, H.
            and Uchida, S. and Davis, J. C.},
  title = {Relating Atomic-Scale Electronic Phenomena
           to Wave-Like Quasiparticle States in
           Superconducting
           {Bi$_2$Sr$_2$CaCu$_2$O$_{8+\delta}$}},
  journal = {Nature},
  volume = {422},
  pages = {592--596},
  year = {2003},
  doi = {10.1038/nature01496}
}

@article{Hardy1993,
  author = {Hardy, W. N. and Bonn, D. A. and Morgan, D. C. and
            Liang, R. and Zhang, K.},
  title = {Precision Measurements of the Temperature Dependence
           of {$\lambda$} in {YBa$_2$Cu$_3$O$_{6.95}$}:
           Strong Evidence for Nodes in the Gap Function},
  journal = {Phys. Rev. Lett.},
  volume = {70},
  pages = {3999},
  year = {1993},
  doi = {10.1103/PhysRevLett.70.3999}
}

@article{Shen1993,
  author = {Shen, Z.-X. and Dessau, D. S. and Wells, B. O. and
            King, D. M. and Spicer, W. E. and Arko, A. J. and
            Marshall, D. and Lombardo, L. W. and Kapitulnik, A. and
            Dickinson, P. and Doniach, S. and DiCarlo, J. and
            Loeser, T. and Park, C. H.},
  title = {Anomalously Large Gap Anisotropy in the
           {$a$-$b$} Plane of
           {Bi$_2$Sr$_2$CaCu$_2$O$_{8+\delta}$}},
  journal = {Phys. Rev. Lett.},
  volume = {70},
  pages = {1553},
  year = {1993},
  doi = {10.1103/PhysRevLett.70.1553}
}

@article{Ding1996,
  author = {Ding, H. and Norman, M. R. and Campuzano, J. C. and
            Randeria, M. and Bellman, A. F. and Yokoya, T. and
            Takahashi, T. and Mochiku, T. and Kadowaki, K.},
  title = {Angle-Resolved Photoemission Spectroscopy Study of
           the Superconducting Gap Anisotropy in
           {Bi$_2$Sr$_2$CaCu$_2$O$_{8+x}$}},
  journal = {Phys. Rev. B},
  volume = {54},
  pages = {R9678},
  year = {1996},
  doi = {10.1103/PhysRevB.54.R9678}
}

@article{Wollman1993,
  author = {Wollman, D. A. and Van Harlingen, D. J. and
            Lee, W. C. and Ginsberg, D. M. and Leggett, A. J.},
  title = {Experimental Determination of the Superconducting
           Pairing State in {YBCO} from the Phase Coherence
           of {YBCO-Pb} dc {SQUIDs}},
  journal = {Phys. Rev. Lett.},
  volume = {71},
  pages = {2134},
  year = {1993},
  doi = {10.1103/PhysRevLett.71.2134}
}

@article{Tsuei1994,
  author = {Tsuei, C. C. and Kirtley, J. R. and Chi, C. C. and
            {Lock See Yu-Jahnes} and Gupta, A. and Shaw, T. and
            Sun, J. Z. and Ketchen, M. B.},
  title = {Pairing Symmetry and Flux Quantization in a
           Tricrystal Superconducting Ring of
           {YBa$_2$Cu$_3$O$_{7-\delta}$}},
  journal = {Phys. Rev. Lett.},
  volume = {73},
  pages = {593},
  year = {1994},
  doi = {10.1103/PhysRevLett.73.593}
}

@article{Tsuei2000,
  author = {Tsuei, C. C. and Kirtley, J. R.},
  title = {Pairing Symmetry in Cuprate Superconductors},
  journal = {Rev. Mod. Phys.},
  volume = {72},
  pages = {969},
  year = {2000},
  doi = {10.1103/RevModPhys.72.969}
}

@article{Kimura2008,
  author = {Kimura, Hikari and Barber, Jr., R. P. and
            Ono, S. and Ando, Yoichi and Dynes, R. C.},
  title = {Scanning Josephson Tunneling Microscopy of
           Single-Crystal
           {Bi$_2$Sr$_2$CaCu$_2$O$_{8+\delta}$}
           with a Conventional Superconducting Tip},
  journal = {Phys. Rev. Lett.},
  volume = {101},
  pages = {037002},
  year = {2008},
  doi = {10.1103/PhysRevLett.101.037002}
}

@article{Hamidian2016,
  author = {Hamidian, M. H. and Edkins, S. D. and
            Joo, Sang Hyun and Kostin, A. and Eisaki, H. and
            Uchida, S. and Lawler, M. J. and Kim, E.-A. and
            Mackenzie, A. P. and Fujita, K. and Lee, Jinho and
            Davis, J. C. S{\'e}amus},
  title = {Detection of a Cooper-Pair Density Wave in
           {Bi$_2$Sr$_2$CaCu$_2$O$_{8+x}$}},
  journal = {Nature},
  volume = {532},
  pages = {343--347},
  year = {2016},
  doi = {10.1038/nature17411}
}

@article{PbSubstitution2019,
  author = {Liu, Jing and Zhao, Lin and Gao, Qiang and
            Ai, Ping and Zhang, Lu and Xie, Tao and
            Huang, Jian-Wei and Ding, Ying and Hu, Cheng and
            Yan, Hong-Tao and Song, Chun-Yao and Xu, Yu and
            Li, Cong and Cai, Yong-Qing and Rong, Hong-Tao and
            Wu, Ding-Song and Liu, Guo-Dong and Wang, Qing-Yan and
            Huang, Yuan and Zhang, Feng-Feng and Yang, Feng and
            Peng, Qin-Jun and Li, Shi-Liang and Yang, Huai-Xin and
            Li, Jian-Qi and Xu, Zu-Yan and Zhou, Xing-Jiang},
  title = {Evolution of Incommensurate Superstructure and
           Electronic Structure with {Pb} Substitution in
           {(Bi$_{2-x}$Pb$_x$)Sr$_2$CaCu$_2$O$_{8+\delta}$}
           Superconductors},
  journal = {Chin. Phys. B},
  volume = {28},
  pages = {077403},
  year = {2019},
  doi = {10.1088/1674-1056/28/7/077403}
}

@article{Drozdov2018,
	author = {Drozdov, I. K. and Pletikosi{\'c}, I. and Kim, C.-K. and
	Fujita, K. and Gu, G. D. and Davis, J. C. S{\'e}amus and
	Johnson, P. D. and Bo{\v z}ovi{\'c}, I. and Valla, T.},
	title = {Phase diagram of {Bi$_2$Sr$_2$CaCu$_2$O$_{8+\delta}$} revisited},
	journal = {Nature Communications},
	volume = {9},
	pages = {5210},
	year = {2018},
	doi = {10.1038/s41467-018-07686-w}
}

@article{Choubey2017,
	author = {Choubey, Peayush and Kreisel, Andreas and Berlijn, T. and
	Andersen, Brian M. and Hirschfeld, P. J.},
	title = {Universality of scanning tunneling microscopy in cuprate superconductors},
	journal = {Phys. Rev. B},
	volume = {96},
	pages = {174523},
	year = {2017},
	doi = {10.1103/PhysRevB.96.174523}
}

\end{document}